\documentclass[preprint2]{aastex}

\usepackage{graphicx}

\def\beq{\begin{equation}}
\def\eeq{\end{equation}}
\def\een#1{\label{#1} \end{equation}}
\def\beqa{\begin{eqnarray}}
\def\eeqa{\end{eqnarray}}
\def\ean#1{\label{#1} \end{eqnarray}}
\def\pd#1#2{\frac{\partial{#1}}{\partial{#2}}}
\def\od#1#2{\frac{d{#1}}{d{#2}}}
\def\eqref#1{(\ref{#1})}
\def\nnn{\nonumber \\}
\def\eps{\varepsilon}

\def\sech{{\rm sech}}

\begin{document}

\title{Note on the single-shock solutions of the
    Korteweg-de Vries-Burgers equation}

\author{Ioannis Kourakis\altaffilmark{1},
    Sharmin Sultana\altaffilmark{1} and
    Frank Verheest\altaffilmark{2,3}}
\altaffiltext{1}{Centre for Plasma Physics, Department of Physics
    and Astronomy, Queen's University Belfast, BT7 1NN
    Northern Ireland, UK}
\altaffiltext{2}{Sterrenkundig Observatorium, Universiteit Gent,
    Krijgs\-laan 281, B--9000 Gent, Belgium}
\altaffiltext{3}{School of Physics, University of KwaZulu-Natal,
    Private Bag X54001, Durban 4000, South Africa}
\email{frank.verheest@ugent.be}


\begin{abstract}

The well-known shock solutions of the Korteweg-de Vries-Burgers equation
are revisited, together with their limitations in the context of plasma
(astro)physical applications.
Although available in the literature for a long time, it seems to have
been forgotten in recent papers that such shocks are monotonic and unique,
for a given plasma configuration, and cannot show oscillatory or
bell-shaped features.
This uniqueness is contrasted to solitary wave solutions of the two parent
equations (Korteweg-de Vries and Burgers), which form a family of curves
parameterized by the excess velocity over the linear phase speed.
\end{abstract}

\maketitle

Among the paradigm nonlinear evolution equations cropping up in various
domains of physics, the Korteweg-de Vries-Burgers (KdVB) equation,
\beq
\pd{\varphi_1}{\tau} + A \varphi_1\, \pd{\varphi_1}{\xi}
    + B \pd{^3 \varphi_1}{\xi^3} = C \pd{^2 \varphi_1}{\xi^2},
\een{kdvb}
arises in physical media where nonlinearity, dispersion and damping
interact on slow timescales to produce solitary structures.
More specifically, in plasma physics \eqref{kdvb} typically obtains by
reductive perturbation analysis of a multi-fluid model, through the use of
coordinate stretching
\beq
\xi = \eps^{1/2}(x - \lambda t), \qquad \tau = \eps^{3/2} t,
\eeq
combined with expansions of the dependent variables like
\beq
\varphi = \eps \varphi_1 + \eps^2 \varphi_2 + \ldots
\eeq
in addition to an appropriate scaling of the damping coefficient, in many
cases due to viscosity.
Here $x$ and $t$ are the original space and time coordinates,
respectively, and $\varphi$ refers to the electrostatic potential of the
solitary waves.
In the absence of damping ($C=0$), the KdVB equation \eqref{kdvb} reduces
to the KdV equation, whereas in the absence of dispersion ($B=0$), it
recovers the Burgers equation, which bears kink-shaped monotonic shock
profile solutions.
All this is well known and has been in the literature for a long time, but
we will have to come back to these points later.

For a purely mathematical study of the properties of the KdVB equation,
\eqref{kdvb} is given and its coefficients $A$, $B$ and $C$ might be
regarded as free parameters.
However, the moment the KdVB equation is derived for a particular plasma
(astro)physical configuration, the precise and often elaborate form of
$A$, $B$ and $C$ has to be computed.
Although the intermediate details need not concern us here, we still have
to remind ourselves that $A$, $B$ and $C$ are functions of the plasma
compositional parameters, which also determine the linear phase velocity
$\lambda$, and thus cannot be chosen randomly.
Moreover, in the process of deriving \eqref{kdvb} one has imposed/used
that $\varphi_1$ vanishes in the undisturbed medium, upstream of the shock
or soliton solutions, translated as $\varphi_1 \rightarrow 0$ for
$\xi \rightarrow +\infty$.
All this has important consequences for the discussion which follows.

Once this is properly kept in mind, there are several ways of deriving the
stationary shock structure of \eqref{kdvb}, by changing to a co-moving
frame with coordinate
\beq
\chi=\kappa(\xi-V\tau),
\een{chi}
where $\kappa$ and $V$ are related to the inverse width and the speed of
the shock, respectively.
Therefore, it is assumed that both $\kappa$ and $V$ are positive.
The shock solutions of the KdVB equation have been in the literature for
a long time, and later rederived by the so-called ``tanh" method,
formalized by \citet{MH1,MH2}.

However, we have to come back in explicit detail to the shock solution of
the KdVB equation, in view of recent misunderstandings about its validity
and its applications, as shown below.
One also has to remember that for all solitary waves, for which explicit
analytical expressions have been obtained, amplitude, width (inversely
related to $\kappa$) and velocity $V$ are inherently linked.
Usually, fixing one of these parameters determines the others.

Now, when looking at several papers in the recent literature
\citep{Shah2009,Saeed2010,Pakzad2011a,Pakzad2011b,Pakzad2011c,%%
Pakzad2011d,PakJav2011,Shah2011,ShahHaqMah}, one sees that $\kappa=1$ is
taken, whether explicitly stated
\citep{Shah2009,Saeed2010,Shah2011,ShahHaqMah}
or only implicitly
\citep{Pakzad2011a,Pakzad2011b,Pakzad2011c,Pakzad2011d,PakJav2011}, by
using the shock solution in the form given by \citet{Shah2009}.
No justification at all is given as to why one would be allowed to put
$\kappa=1$, nor is there any discussion of the consequences.
As we will see, taking $\kappa=1$ is not only needlessly stringent, but
also erroneous, and in many cases one is not even able to verify that it
holds, given the complexities in the expressions for $A$, $B$ and $C$,
except for specific numerical choice of all plasma parameters.
Some other papers even leave $\kappa$ undetermined, as if it were a free
parameter \citep{Mahm,Akhtar}.

When the transformation \eqref{chi} is applied to \eqref{kdvb}, one finds
\beq
- \kappa V\, \od{\varphi_1}{\chi}
    + A \kappa \varphi_1\, \od{\varphi_1}{\chi}
    + B \kappa^3 \od{^3 \varphi_1}{\chi^3}
    - C \kappa^2 \od{^2 \varphi_1}{\chi^2} = 0.
\een{kdvbN}

One of the popular methods of finding the shock structure for \eqref{kdvb}
is through the tanh method, and we will follow the original paper by
\citet{MH1}, rather than a vast array of newcomers.
We are forced to do so, to point out where the specific restrictions to
plasma (astro)physics applications play a role and to correct some uses in
the literature which have strayed in this respect from the original
solutions \citep{MH1} already available.
Our treatment here is more general than that of \citet{MH1}, because in
their paper $A=1$ has been taken.
While one can always rescale the absolute value of some of the
coefficients in \eqref{kdvb}, one cannot easily do away with the sign, and
we keep therefore $A$ as determined by the plasma model under
consideration.

Using the transformation $\alpha = \tanh\chi$ in \eqref{kdvbN} and noting
that $d\alpha/d\chi = 1-\tanh^2\chi$, we obtain
\beqa
&& - V\, \od{\varphi_1}{\alpha}
+ A \varphi_1\, \od{\varphi_1}{\alpha} \nnn
&& \qquad +\, B \kappa^2 \od{}{\alpha} \left\{ (1-\alpha^2) \od{}{\alpha}
    \left[ (1-\alpha^2) \od{\varphi_1}{\alpha} \right] \right\} \nnn
&& \qquad -\, C \kappa \od{}{\alpha}
    \left[ (1-\alpha^2)\od{\varphi_1}{\alpha} \right] = 0.
\ean{kdvbT}
Here one common factor $\kappa$ and one common bracket $(1-\alpha^2)$ have
already been divided out, to simplify the subsequent computations.

The idea is then to look for solutions $\varphi_1$ as a finite power
series in $\alpha$, which in this case (and in many others) will end with
the quadratic term \citep{MH1}, thus
\beq
\varphi_1 = \beta_0 + \beta_1 \alpha + \beta_2 \alpha^2.
\een{alpha}
The reason that the power series breaks off comes from a balance between
the highest nonlinearity and dispersive terms in \eqref{kdvbT}.
Given that the different powers of $\alpha$ are functionally independent,
we get a system of algebraic equations,
\beqa
\alpha^0: && -V\beta_1 + A\beta_0\beta_1 - 2B\kappa^2\beta_1
    - 2C\kappa\beta_2 = 0, \label{a0} \\
\alpha^1: && -2V\beta_2 + 2A\beta_0\beta_2 + A\beta_1^2
    - 16B\kappa^2\beta_2 \nnn
  && \qquad +\, 2C\kappa\beta_1 = 0, \label{a1} \\
\alpha^2: && 3A\beta_1\beta_2 + 6B\kappa^2\beta_1
    + 6C\kappa\beta_2 = 0, \label{a2} \\
\alpha^3: &\quad& 2A\beta_2^2 + 24B\kappa^2\beta_2 = 0, \label{a3}
\eeqa
determining the as yet unknown coefficients $\beta_0$, $\beta_1$ and
$\beta_2$.
Solve first \eqref{a3} for $\beta_2$ to find
\beq
\beta_2 = -\, \frac{12B\kappa^2}{A},
\een{b2}
and substitute this in \eqref{a2}.
This allows now to obtain
\beq
\beta_1 = -\, \frac{12C\kappa}{5A}.
\een{b1}
Solving next \eqref{a0} yields
\beq
\beta_0 = \frac{V}{A} + \frac{12B\kappa^2}{A}.
\een{b0}
Although all coefficients needed for \eqref{alpha} have now been
determined, there is still one condition to be satisfied before the scheme
can work, namely \eqref{a1}.
This was apparently overlooked or not deemed important
\citep{Shah2009,Saeed2010,Mahm,Akhtar,Pakzad2011b,Shah2011,ShahHaqMah,%%
Pakzad2011b}, while others
\citep{Pakzad2011a,Pakzad2011c,Pakzad2011d,PakJav2011} just copied the
erroneous solution, without going through the algebra.
Working out \eqref{a1}, one arrives at
\beq
\kappa = \frac{C}{10B},
\een{k}
where for simplicity we have taken both $B$ and $C$ positive, as they
usually are in most examples found in the literature.
Adopting other sign conventions can easily be incorporated but would add
nothing to the physics.
Indeed, it is straightforward to see that minus signs can be handled in
the general solution by appropriate space and/or time reversals.
Note in passing that
$\kappa\tanh[\kappa(\xi-V\tau)] = -\kappa\tanh[-\kappa(\xi-V\tau)]$, for
any real $\kappa$.

At this stage it is clear how serious a restriction $\kappa=1$ is, for two
separate reasons.
First, all solitary wave characteristics show an inherent link between
amplitude, width (inversely related to $\kappa$) and velocity $V$ of the
structure, and arbitrarily fixing one narrows the choices enormously.
Second, assuming $\kappa=1$ means from \eqref{k} that $C=10 B$, a relation
which usually cannot be obeyed by inserting some numbers in the rather
complicated expressions $B$ and $C$, as a glance at the papers involved
\citep{Shah2009,Mahm,Saeed2010,Akhtar,Pakzad2011a,Pakzad2011b,%%
Pakzad2011c,Pakzad2011d,PakJav2011,Shah2011,ShahHaqMah}
will immediately reveal.
Taken together, this implies that the resulting numerics, graphs and
discussions
\citep{Shah2009,Mahm,Saeed2010,Akhtar,Pakzad2011a,Pakzad2011b,%%
Pakzad2011c,Pakzad2011d,PakJav2011,Shah2011,ShahHaqMah} cannot be trusted.

Using now \eqref{k} in the coefficients \eqref{b2}--\eqref{b0} shows that
\beqa
&& \beta_0 = \frac{V}{A} + \frac{3C^2}{25AB}, \quad
\beta_1 = -\, \frac{6C^2}{25AB}, \nnn
&& \beta_2 = -\, \frac{3C^2}{25AB}.
\ean{b012}
At this stage the shock solution is
\beq
\varphi_1 = \frac{3C^2}{25AB}\left(1-\tanh^2\chi\right) + \frac{V}{A}
    - \frac{6C^2}{25AB}\tanh\chi.
\eeq
Since $B$ and $C$ are assumed positive, it is the sign of $A$ which will
be determining the polarity of the kink solution.
However, this only obeys the requirement that $\varphi_1 \rightarrow 0$
for $\xi \rightarrow +\infty$ provided one takes
\beq
V = \frac{6C^2}{25B} = 24B\kappa^2.
\een{V}
Also this inherent aspect of the correct solution has been overlooked in
some of the recent papers \citep{Shah2009,Pakzad2011a,Pakzad2011b,%%
Pakzad2011c,Pakzad2011d,PakJav2011}.
The second expression for $V$ in \eqref{V} clearly shows the link between
width (through $\kappa$) and velocity of the structure, and for right
propagating structures $V$ is taken positive, which therefore requires
$B$ to be positive.

\begin{figure}
\includegraphics[width=80mm]{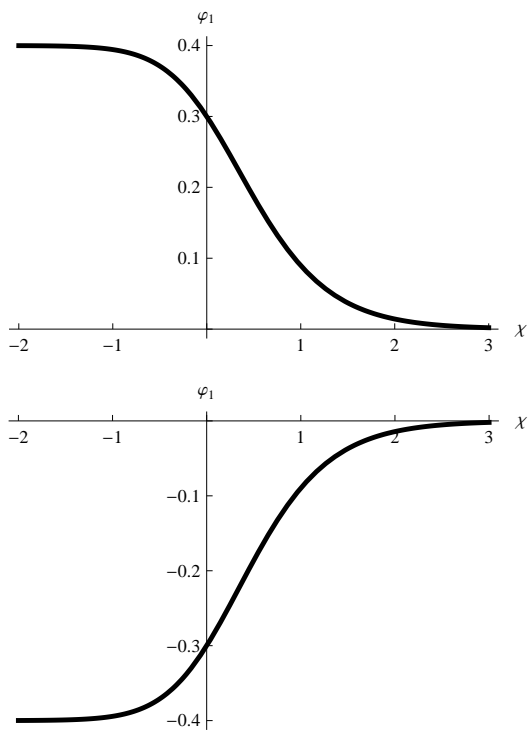}
\caption{
Typical KdVB shock profile, where the amplitude $3C^2/(25AB)=0.1$ has
been taken for the upper panel and $-0.1$ for the lower panel.}
\label{shock}
\end{figure}

Finally, we arrive at the shock solution as
\beq
\varphi_1 = \frac{3C^2}{25AB} \left[1-\tanh^2\chi + 2(1-\tanh\chi)\right],
\een{sol}
where in $\chi$ we have to insert \eqref{k} and \eqref{V}, giving
\beq
\chi = \frac{C}{10B} \left(\xi - \frac{6C^2}{25B}\, \tau \right).
\een{chiN}
The kink structure \eqref{sol} is unique, since for a given plasma
configuration the compositional parameters fully determine $A$, $B$ and
$C$, and hence there is one and only one shock solution, the generic
profile of which we illustrate in Fig.\ \ref{shock}, once for a positive
(upper panel), once for a negative (lower panel) polarity.
This point has already been made before \citep{MH1}, in a mathematical
discussion, almost in passing, without really stressing its consequences
for detailed plasma (astro)physics problems.

Further remarks are in order here.
Since \eqref{sol} can be rewritten as
\beq
\varphi_1 = \frac{3C^2}{25AB} \left[4-(1+\tanh\chi)^2\right],
\een{solK}
the kink is always monotonic, and no oscillatory part nor peak or
bell-shaped curve may appear in its graph, contrary to what is found in
recent papers
\citep{Shah2009,Mahm,Saeed2010,Akhtar,Pakzad2011a,Pakzad2011b,%%
Pakzad2011c,Pakzad2011d,PakJav2011,ShahHaqMah}.
There may be physical situations where shocks including oscillatory
trails or precursors are observed, but these cannot be described by the
KdVB formalism.

Note that when $C=0$, the whole shock structure disappears.
This is a direct consequence of the very delicate balance needed between
a solitary wave (KdV) and a shock wave (Burgers) to form the combined
solution \citep{MH1}.
To see this more explicitly, substitute in \eqref{sol}
$1-\tanh^2\chi=\sech^2\chi$, which is reminiscent of the typical KdV
one-soliton solution.
In addition, since reductive perturbation analysis requires that
$\varphi_1$ be small enough to neglect higher-order effects,
$3C^2/(25|AB|)$ should be rather smaller than 1.

All this has to be contrasted to what happens when $C=0$ and \eqref{kdvb}
reduces to the standard KdV equation, without dissipation through
viscosity, or when $B=0$ and \eqref{kdvb} becomes the Burgers equation,
in the absence of dispersion.
Furthermore, when $C=0$ the KdV $\sech^2\chi$ soliton cannot be
directly recovered, contrary to what is claimed in the literature
\citep{Shah2009,Saeed2010,Pakzad2011a,Pakzad2011b,Pakzad2011c,%%
Pakzad2011d,PakJav2011,Shah2011,ShahHaqMah}.

To see the differences, let us now first put $C=0$, return to
\eqref{a0}--\eqref{a3} and go again through the motions.
It turns out that $\beta_2$ is still given by \eqref{b2}, but $\beta_1=0$
and \eqref{b0} is replaced here by
\beq
\beta_0 = \frac{V}{A} + \frac{8B\kappa^2}{A}.
\een{b00}
Hence, to arrive at the typical KdV soliton solution in
$\sech^2\xi = 1 - \tanh^2\xi$, obeying $\varphi_1 \rightarrow 0$ when
$\xi \rightarrow \pm\infty$, it is required that
\beq
V = 4B\kappa^2,
\een{VV}
and now for each superacoustic soliton velocity $V$ one finds a soliton of
the form
\beq
\varphi_1 = \frac{3V}{A}
    \sech^2\left[ \frac{1}{2}\, \sqrt{\frac{V}{B}}\, ( \xi - V\tau)\right].
\een{kdv}
Here $B>0$ is needed, which is usually the case, and the soliton polarity
is given by the sign of $A$.

Doing a similar exercise for the Burgers equation, with $B=0$, leads from
\eqref{a2} and \eqref{a3} to $\beta_2=0$, in other words, \eqref{alpha}
stops at the linear term \citep{MH1}.
Now \eqref{a0} and \eqref{a1} give that
\beq
\beta_0 = \frac{V}{A}, \qquad \beta_1 = - \frac{2C\kappa}{A},
\eeq
and the proper solution needs
\beq
V = 2C\kappa.
\eeq
Taking again $V$ as the free parameter, the shock solution is found as
\beq
\varphi_1 = \frac{V}{A} \left\{ 1
    - \tanh\left[ \frac{V}{2C} (\xi - V\tau) \right]\right\}.
\een{burg}
With the appropriate changes of notation, the solutions \eqref{kdv} and
\eqref{burg} can be found in the original discussion by \citet{MH1}.

To conclude, we have discussed the intricacies of the proper derivation of
the solitary shock structure and its limitations in the context of plasma
(astro)physical applications.
Although these results and restrictions have been in the literature for a
long time \citep{MH1,MH2}, it seems to have been forgotten in recent
papers \citep{Shah2009,Mahm,Saeed2010,Akhtar,Pakzad2011a,Pakzad2011b,%%
Pakzad2011c,Pakzad2011d,PakJav2011,Shah2011,ShahHaqMah} that a shock
modeled by \eqref{sol} can only be monotonic, without oscillations or
peaks, and is, moreover, unique.

This also holds for the coefficients $A$, $B$ and $C$, once specific
numbers have been assigned to the various compositional parameters in the
plasma model under consideration, and therefore $A$, $B$ and $C$ cannot be
treated as free parameters, as they might be in a purely mathematical
discussion of the properties of \eqref{kdvb}.
But even then they determine $V$ and $\kappa$ in a unique way.

One sees that the solitary wave solutions of the two parent nonlinear
equations, the KdV and the Burgers equations, are different in character,
as they form one-parameter families of curves, dependent on the free
choice of the excess velocity $V$ above the linear phase speed $\lambda$.

\acknowledgments
I.K. and S.S. acknowledge funding from the UK EPSRC (Engineering and
Physical Science Research Council) via a Science and Innovation award to
Centre for Plasma Physics, Queen's University Belfast (grant no.\
EP/D06337X/1).

%%%%%%%%%%%%%%%%%%%%%%%%%

\end{document}